# Behaviour and Perception-based Pedestrian Evacuation Simulation

**Tobias Kretz[1a], Georg Mayer[1b], and Andreas Mühlberger[2]**

[1]PTV Planung Transport Verkehr AG,
[1a]Stumpfstr. 1, D-76131 Karlsruhe
[1b]Kriegerstraße 15, D-70191 Stuttgart

[2]Psychologie I, Universität Würzburg,
  Marcusstraße 9-11, D-97070 Würzburg

Corresponding author: georg.mayer@ptv.de

**Abstract**   This contribution reports on the research project SKRIBT and some of its results. An evacuation simulation based on VISSIM's pedestrian dynamics simulation was developed, that - with high time resolution - integrates results from studies on behavior in stress and crisis situations, results from CFD models for e.g. fire dynamics simulations, and considers visibility of signage and — adding a psychological model — its cognition. A crucial issue is the cognition of smoke or fire by the occupant and his / her resulting spontaneous or deliberate reaction to this episode.

## Introduction and Motivation

To estimate numbers of affected occupants in cases of emergency evacuations, simulation models have been developed that already for a number of years not only perform microscopic simulations of pedestrian dynamics, but also calculate effects from physical hazards on the occupants — mostly by means of fractional effective dose models. But to our knowledge there is no model that considers perception and behavior in such special environments more than to a rudimentary degree.

As part of the research project SKRIBT [1] an evacuation simulation based on VISSIM's pedestrian dynamics simulation [2] was developed, that - with high time resolution - integrates results from studies on behavior in stress and crisis situations, results from CFD models for e.g. fire dynamics simulations, and considers visibility of signage and — adding a psychological model — its cognition. A crucial issue is the cognition of smoke or fire by the occupant and his / her resulting spontaneous or deliberate reaction to this episode.

## Simulation of Pedestrians

For the calculation of damage degree and countermeasure efficacy one needs models with high resolution in time and space to determine the influences on the occupants. This was done using numerical calculation schemes (CFD models).

The simulation of pedestrian dynamics was done with VISSIM, using the Social Force Model [3]. The latest version of the model [4] was extended, so it can be applied for the simulation of conflict areas (interaction of pedestrians and vehicles), multi-storey buildings, pedestrians as passengers of public transport, and waiting queues.

Route choice and pre-movement time depend for one on a calculation of visibility of exit signage (and the routes that are linked with a sign) and after by the principal availability of information that has been calculated by a behavioral model. The successful reaction on external influences requires both the perception and the cognition of the situation and information. Apart from individual parameters, this depends on the local conditions.

The calculation of visibility considers the orientation of an occupant in space, his body height, obstacles like trucks standing still, smoke concentration along the line of sight, and a sign's luminosity. Not considered are occlusion by moving objects like other occupants or vehicles. This calculation is rather straight forward, as it is based on long-known and well-defined physical and geometrical laws and states. The only biological and therefore individual constant entering is the threshold at which a sign can be distinguished from its background.

The behavioral and decision model contrary to that has been developed newly in cooperation with the University of Würzburg[1]. It is based on a factor concept that adds and weighs different factors influencing a decision to start evacuation or the ability to become aware of a sign that is visible to a specific occupant. Additionally walking speed is influenced by visibility conditions [5].

The cognition of smoke (resp. fire) by the occupants mostly takes place before any instructions by the tunnel operators or rescue services are given. The model calculates the amount of psychical stress the occupant suffers from, depending on internal (individuals´ state, fearfulness, behavioral routines, knowledge, …) and external parameters (intensity of danger, social influence, …), by means of a "potential for reaction". It triggers a decision progress to flee, that leads to either an impulsive or reflective action. Both types are distinguished by the model, and the resulting paths of action considered.

Finally, to determine the damage amount on individual occupants the Fractional Effective Dose (FED) model by Purser is applied [6].

---

[1] Paul Pauli, Andreas Mühlberger, Silke Eder, Johanna Brütting, University of Würzburg, Department of Psychology, Biological Psychology, Clinical Psychology and Psychotherapy, Marcusstr. 9-11, D-97070 Würzburg, Germany

## Application and Results

The system has been tested extensively and, as a SKRIBT project, been applied to a tunnel incident scenario. There routing is simple, as there is only one fully developed dimension. Occupants need to decide when to start moving and can either head towards the tunnel portal or recognize and use emergency exits.

In this example, two 1,200 m long tunnels have been modeled. The first one has a constant gradient of 3 %, the second one a de- and an ascending leg. The incident takes place in the middle of the tunnel. Emergency exits are 300 m apart.

So far, the scenarios "tunnel fire" - due to the discharge and ignition of fuel or propane - and "toxic gas" - modeling the discharge of chlorine or ammonia - have been analyzed. In all cases, both a spontaneous discharge with major consequences as well as a continuous discharge with minor effects have been applied.

Figure 1 shows a tunnel fire (burning fuel of a road tanker). It is apparent how people recognize the incident either by direct view of the flames or the cognition of smoke, which spreads with a velocity of approx. 7 m/s, due to the stack effect. People start to flee, triggering other people still in their vehicles to do the same. They are hindered by diminished visibility and breathing problems. It is assumed that detection and general alarm, takes place 60 s after the incident. It is assumed that then all people flee. After about 60 s, people begin to queue up at the emergency exits. Fatal casualties and unconscious persons are depicted in red.

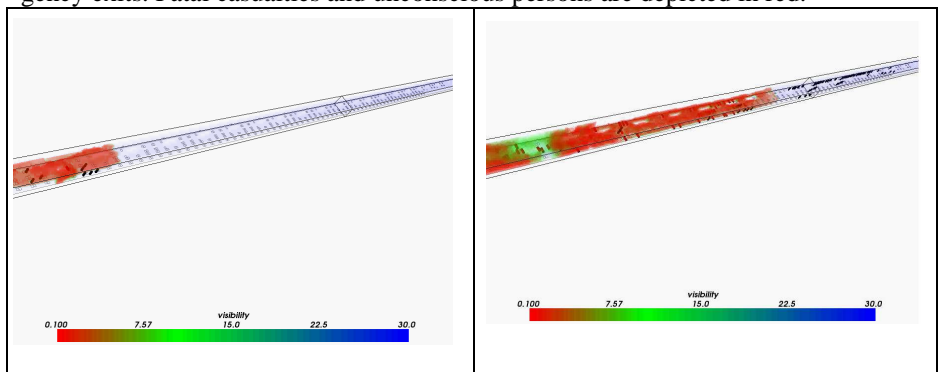

**Figure 1: Tunnel fire (fuel) after 5 s (left) and 60 s (right)**

Figure 2 shows the spontaneous discharge of 21 t chlorine. Chlorine appears in high concentrations as yellow-green fumes. In tunnel lighting conditions, these are not easily discernible. Therefore, it is assumed that only people nearest to the incident – those who are able to see that a vehicle transporting hazardous goods leaking some kind of substance is involved – start to flee. These trigger others to follow, but the process takes more time than in the first example. The effects of inhaling chlorine are more severe than for smoke. On the other hand, visibility – affecting orientation as well as perceptibility of emergency signs and exits – is not impaired as it would be by smoke, and the distribution of the gas is slower. As no heat is involved, no incident detection takes place in this scenario. In the end, the resulting number of casualties is about the same as in the first example.

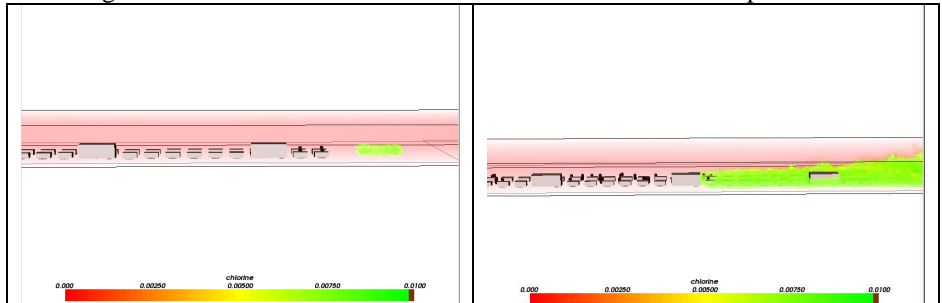

**Figure 2: Spontaneous discharge of 21 t chlorine after 5 s (l.) and 60 s (r.)**

## Summary

The model allows the realistic reproduction of episodes and respective escape situations. Its main applications will be the determination of extents of loss within risk analysis and the development of escape and evacuation concepts.

## Acknowledgements

SKRIBT is funded by the German Ministry of Education and Science (BMBF). The authors thank Dmitri Danewitz and Peter Ehrhardt for SKRIBT-related project work and software development.